\documentclass[jgr,draft]{agujournal2019}
\usepackage{apacite}
\usepackage{url} 
\usepackage{lineno}
\usepackage{soul}
\usepackage{multirow}
\usepackage{graphicx}
\usepackage{colortbl}
\usepackage{color}
\usepackage{amssymb}
\usepackage{amsmath}
\usepackage{caption}

\journalname{JGR: Space Physics}

\begin{document}

%
%


\title{Multi-scale coupling during magnetopause reconnection: interface between the electron and ion diffusion regions}

%
%




\authors{	K. J. Genestreti\affil{1},
		Y.-H. Liu\affil{2},
	      	T.-D. Phan\affil{3},
		R. E. Denton\affil{2},
		R. B. Torbert\affil{1,4},
		J. L. Burch\affil{5},
		J. M. Webster\affil{6,5},
		S. Wang\affil{7,8},
		K. J. Trattner\affil{9},
		M. R. Argall\affil{4},
		L.-J. Chen\affil{7},
		S. A. Fuselier\affil{5,10},
		N. Ahmadi\affil{9},
		R. E. Ergun\affil{9},
		B. L. Giles\affil{7},
		C. T. Russell\affil{11},
		R. J. Strangeway\affil{11},
		S. Eriksson\affil{9}}
		
\affiliation{1}{Southwest Research Institute, Durham, New Hampshire, USA}
\affiliation{2}{Dartmouth College, Hanover, New Hampshire, USA}
\affiliation{3}{University of California Berkeley, Berkeley, California, USA}
\affiliation{4}{University of New Hampshire, Durham, New Hampshire, USA}
\affiliation{5}{Southwest Research Institute, San Antonio, Texas, USA}
\affiliation{6}{Rice University, Houston, Texas, USA}
\affiliation{7}{NASA Goddard Space Flight Center, Greenbelt, Maryland, USA}
\affiliation{8}{University of Maryland, College Park, Maryland, USA}
\affiliation{9}{University of Colorado Boulder, Boulder, Colorado, USA}
\affiliation{10}{University of Texas San Antonio, San Antonio, USA}
\affiliation{11}{University of California Los Angeles, Los Angeles, California, USA}





\correspondingauthor{Kevin J. Genestreti}{kevin.genestreti@swri.org}




\begin{keypoints}
\item With ion-scale inter-spacecraft separations Magnetospheric Multiscale crossed an ion diffusion region with embedded electron currents 
\item Intense pileup of reconnected magnetic flux is observed within the ion diffusion region where the exhaust opens at a wide angle
\item Kinetic simulations indicate that the intense flux pileup may be a standing feature related to the asymmetric Hall electron currents
\end{keypoints}

%
%


\begin{abstract}
Magnetospheric Multiscale (MMS) encountered the primary low-latitude magnetopause reconnection site when the inter-spacecraft separation exceeded the upstream ion inertial length. Classical signatures of the ion diffusion region (IDR), including a sub-ion-Alfv\'enic de-magnetized ion exhaust, a super-ion-Alfv\'enic magnetized electron exhaust, and Hall electromagnetic fields, are identified. The opening angle between the magnetopause and magnetospheric separatrix is $30^\circ\pm5^\circ$. The exhaust preferentially expands sunward, displacing the magnetosheath. Intense pileup of reconnected magnetic flux occurs between the magnetosheath separatrix and the magnetopause in a narrow channel intermediate between the ion and electron scales. The strength of the pileup (normalized values of 0.3-0.5) is consistent with the large angle at which the magnetopause is inclined relative to the overall reconnection coordinates. MMS-4, which was two ion inertial lengths closer to the X-line than the other three spacecraft, observed intense electron-dominated currents and kinetic-to-electromagnetic-field energy conversion within the pileup. MMS-1, 2, and 3 did not observe the intense currents nor the particle-to-field energy conversion but did observe the pileup, indicating that the edge of the generation region was contained within the tetrahedron. Comparisons with particle-in-cell simulations reveal that the electron currents and large inclination angle of the magnetopause are interconnected features of the asymmetric Hall effect. Between the separatrix and the magnetopause, high-density inflowing magnetosheath electrons brake and turn into the outflow direction, imparting energy to the normal magnetic field and generating the pileup. The findings indicate that electron dynamics are likely an important influence on the magnetic field structure within the ion diffusion region.
\end{abstract}

\section*{Plain Language Summary}
The Earth's and Sun's magnetic fields meet and can interconnect at the outermost boundary of the Earth's magnetosphere, the magnetopause. Reconnection of the two magnetic fields requires that the motions of the ions and electrons become decoupled from the motion of the field itself. Owing to their greater inertia, the ions becomes decoupled from the magnetic field within a much larger volume of space compared to the electrons. The demagnetization of ions and electrons during magnetic field reconnection are difficult to study simultaneously with in-situ data, as both the larger ion and smaller electron scale sizes need to be simultaneously resolved. In this study, we report an observation of magnetopause reconnection with the Magnetospheric Multiscale (MMS) mission. MMS has instruments capable of resolving the electron scales, and we analyze an event for which the spacecraft were separated by a large enough distance to resolve the ion scales. We find that the electron dynamics are important for influencing the structure of the magnetic fields in the region where ion motions are decoupled from the field.

%
%

%


%
%
%
%

\section{Introduction}

The efficiency of magnetic reconnection is thought to be set by the diffusion region and its boundary conditions \cite{CassakandShay.2007}. The diffusion region consists of an ion-kinetic-scale ion diffusion region (IDR) wherein ions are demagnetized, electrons are magnetized, and Hall effects accelerate and broaden the exhaust jets \cite{Sonnerup.1979,Birn.2001,Cassak.2017b}. The IDR encompasses a central electron diffusion region (EDR) \cite{Vasilyunas.1975,Sonnerup.1979,Burch.2016b}, wherein all species are demagnetized and field lines reconnect at an X-point. Super-ion-Alfv\'enic, i.e., faster than the ion Alfv\'en speed, electron jets may extend tens of ion skin depths downstream of the central EDR \cite{Karimabadi.2007,Shay.2007,Phan.2007,Chen.2008}. These extended jets in the so-called outer EDR carry the perpendicular portion of the Hall current system. The jets eventually brake and magnetize \cite{Hwang.2017}, imparting energy to the normal magnetic field. The Hall current is closed by (largely) parallel currents that extend outward from the EDR along the separatrices \cite{Sonnerup.1979}.

A number of theoretical works suggest that IDR processes, rather than EDR processes, typically set the collisionless reconnection rate \cite{Birn.2001,Shay.2001,Drake.2008,Cassak.2017b,Liu.2017,Liu.2018a}. The transition from the narrow whistler-mediated electron outflow to the broad, Alfv\'en-wave-like fluid exhaust is thought to be driven by the dispersive Hall effect \cite{Mandt.1994}. Simulations have determined that the reconnection rate is insensitive to the electron dissipation \cite{Birn.2001}; this follows from fully-kinetic and Hall magnetohydrodynamic (MHD) simulations, which produced practically identical reconnection rates, and the fact that the Hall force does not directly dissipate energy. Aforementioned theoretical works are largely based on two-dimensional simulations of reconnection with steady upstream conditions, a framework that suppresses the growth of many instabilities \cite{Daughton.2014,Price.2016,Price.2017,Le.2017}. Magnetospheric Multiscale (MMS) magnetopause observations demonstrated that electron dynamics in the EDR can locally (in time and space) modify the reconnection electric field \cite{Genestreti.2018a,Burch.2018}. The growth of 3-d current sheet instabilities near X-points is also frequently observed \cite{Ergun.2017,Graham.2019}. The energy conversion rate in the EDR, which is proportional to the overall reconnection rate during laminar reconnection \cite{Genestreti.2018c,Nakamura.2018}, often exhibits large positive and negative fluctuations that are up to orders of magnitude larger than expected. ``Bursty'' electron flows are ubiquitous in magnetopause EDRs \cite{Genestreti.2017,Genestreti.2018a,Cassak.2017a,Burch.2018,Webster.2018}. Given that the electrons accelerated from the EDR form a small segment of the much larger Hall current system, we seek to determine whether fluctuations in the output of the EDR affect the reconnection rate in the larger IDR.

This study investigates a primary magnetopause reconnection site observed by MMS while the spacecraft were separated by distances exceeding the upstream ion inertial scale. The primary X-line is thought of as the dominant site of magnetosphere-magnetosheath reconnection and is distinguished from secondary X-lines that form in the downstream exhaust (see Figure 8 of \citeA{Fuselier.2018}). The overarching goal of the investigation is to analyze the coupling between the EDR processes and the structure of the IDR. Specifically, we analyze the small-scale electron dynamics in the Hall current region and determine their impact on the larger-scale IDR structure. We find that MMS-4, which was two ion inertial lengths closer to the X-line than the other spacecraft, observed electron braking, particle-to-field energy conversion, and perpendicular electron Hall currents. The other three spacecraft did not observe these signatures, which may indicate that the edge of the outer EDR was between MMS-4 and the other three spacecraft. The opening angle of the exhaust was larger on the magnetosheath side of the exhaust ($>24^\circ\pm4^\circ$) than on the magnetospheric side ($6^\circ\pm4^\circ$). Intense and localized pileup of reconnected magnetic flux $B_N$ was observed at the magnetosheath side of the magnetopause by all spacecraft, and $B_N$ was most intense at the locations of MMS-1, 2 and 3, downstream of MMS-4. MMS-4, uniquely, observed the generation region with particle-to-field energy conversion. Comparisons with two particle-in-cell simulations suggest that these features are all interconnected aspects of the asymmetric Hall current, which contributes to the (asymmetric) opening angle of the magnetopause IDR. The pileup is likely related to the magnetosheath electron dynamics; the higher-momentum electrons on the magnetosheath side turn toward then away from the X-line as they cross the separatrix and enter the exhaust, imparting energy to the normal magnetic field where the brake and turn. The larger normal magnetic field on the magnetosheath side of the exhaust leads to the larger opening angle on the magnetosheath side, compared with that of the lower-momentum magnetospheric side. 

The manuscript is laid out as follows: section 2 describes the dataset. Section 3 provides context for the magnetopause crossing and the upstream conditions (Table 1). Section 4 analyzes the ion-scale structure of the IDR, which is used to determine the path of MMS and the structure of the magnetopause (Figure \ref{ions}i). Section 5 determines the opening angle of the exhaust and section 6 analyzes the pileup region of reconnected magnetic flux observed in a thin (thickness in the normal direction is intermediate between ion and electron inertial scales) and elongated (length in the outflow direction is greater than the ion inertial length) channel embedded within the IDR. Section 7 compares the MMS observations with two 2.5-d particle-in-cell (PIC) simulations. Section 8 summarizes and interprets our findings.

\section{Instrumentation and data}

Simultaneous resolution of electron and ion-scale dynamics is required to understand the EDR/IDR interface. MMS \cite{Burch.2016a} has the time resolution required to resolve the electron scale and, for this event, an average inter-spacecraft separation of $\sim$1.5 ion skin depths (73 km), permitting resolution of ion-scale dynamics.

MMS surveyed the low-latitude magnetopause with an apogee of 12 Earth radii ($\mathrm{R_E}$) from 2015--2017. Early in this interval, the inter-spacecraft separations exceeded the typical magnetosheath ion skin depth \cite{Fuselier.2016}. Data from the fast plasma instruments (FPI) \cite{Pollock.2016}, electric field double probes (EDP) \cite{Lindqvist.2016,Ergun.2016a}, and fluxgate magnetometers (FGM) \cite{Russell.2016} are used. Distributions and moments of ions and electrons are obtained by FPI once per 150 ms and 30 ms, respectively. Data from the hot plasma composition analyzer (HPCA) \cite{Young.2016}, which is able to detect colder plasma compared to FPI, are used to determine asymptotic upstream number densities. The 3-d magnetic field is measured by FGM at 128 vectors per second. The 3-d electric field is measured by EDP at 8,196 vectors per second. Level 3 electric field data are used, which are calibrated to remove running offsets from the electron convective field. 

\section{Magnetopause upstream conditions}

\begin{figure}
\noindent\includegraphics[width=39pc]{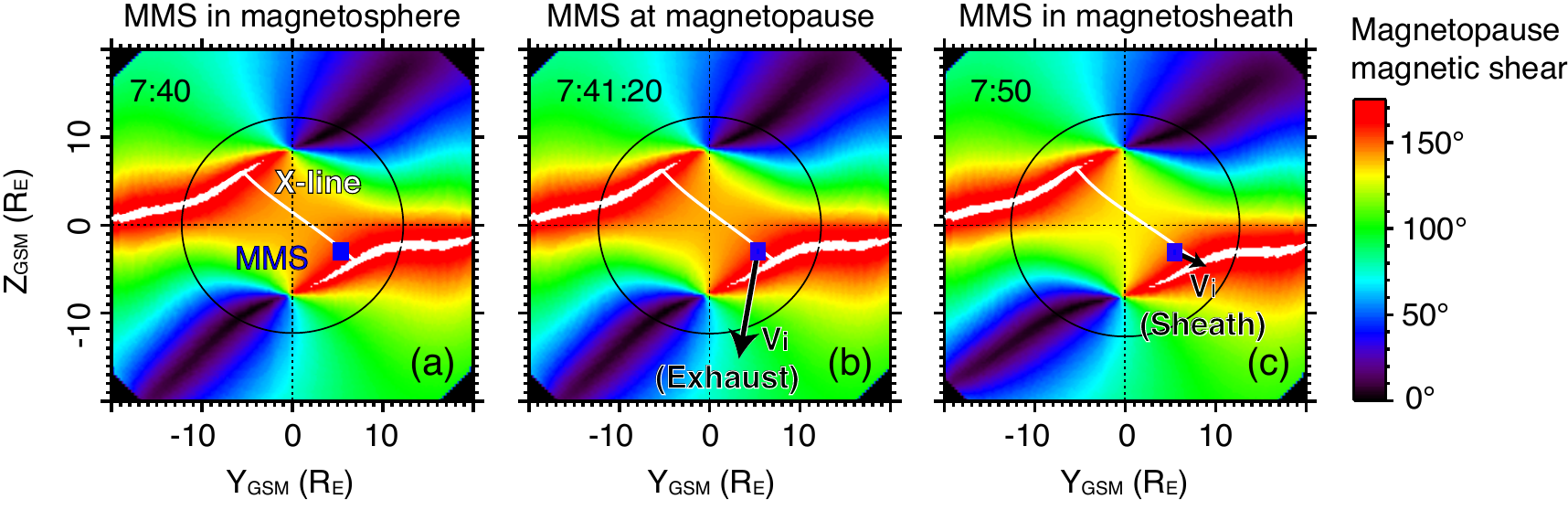}
\caption{The maximum magnetic shear magnetopause model and observed ion flow (a) before, (b) during, and (c) after the 7:41 UT magnetopause crossing. The view is from the sun and the color shows the shear angle between the model-draped WIND magnetic field and the T96 magnetospheric field model \cite{Trattner.2007,Trattner.2016}. The white trace is the predicted X-line location, the blue square is MMS, the black circle is the terminator, and the black arrows are the ion bulk velocity measured by MMS-FPI in the spacecraft velocity frame (note the magnetospheric flow in this frame is too small to be visible).}
\label{msm}
\end{figure}

The MMS magnetopause crossing occurred on 2015 September 19 at 7:41 universal time (UT) at the location shown in Figure \ref{msm}. During $\sim$7 to 10 UT MMS skirted the magnetopause near the predicted location of the primary low-latitude X-line \cite{Trattner.2016}. Many diffusion region or near-diffusion region encounters occurred in this interval \cite{Chen.2016,Trattner.2016,WangS.2016,Wilder.2016,Hwang.2017}, indicating persistent reconnection along the low-latitude magnetopause. \citeA{WangS.2016} investigated ion acceleration during this 7:41 UT crossing and \citeA{Chen.2016} identified an encounter with the central EDR at 7:43 UT (see Section 7.2). Figure \ref{msm} shows (1) that at 7:41 UT MMS was close to the primary X-line location determined by \citeA{Trattner.2016} and (2) that ion outflow (Fig. \ref{msm}b) was observed as MMS transitioned from the comparatively stagnant magnetospheric plasma (\ref{msm}a) to the magnetosheath flow (Fig. \ref{msm}c).

Relevant parameters for the 7:41 UT magnetopause crossing are provided in Table 1. Of particular interest is the hybrid upstream inertial length $d_{i0}=49.3$ km, which is smaller than the average inter-spacecraft separation, 73 km. The local orientation and motion of the magnetopause are determined in supporting information (Appendix A). $LMN$ coordinates are determined by applying joint variance analysis (JVA) \cite{MozerandRetino.2007} to the electric and magnetic field observed by MMS-4. Figure \ref{lmn_vstr}a shows excellent agreement between these $LMN$ coordinates and four independently determined $LMN$ systems. The velocity of the reconnection site, given in Table 1, is strongly southward, moderately duskward, and weakly earthward. The velocity was determined using the spatiotemporal difference technique \cite{Shi.2006,Shi.2019} and is favorably compared with results from timing analysis \cite{ISSIchap10} (see Appendix A). 

\begin{table}
\centering
\caption{Asymptotic upstream conditions and additional parameters determined from MMS-4 data during magnetosphere (7:35--38 UT) and magnetosheath (7:50-8:00 UT) intervals. HPCA data are used to determine the ion parameters in the upstream magnetosphere, as the instrument is uniquely capable of characterizing the cold (few eV) magnetospheric ions. The hybrid reconnecting magnetic field component $B_{L0}$, hybrid ion Alfv\'en speed $V_{Ai0}$, and hybrid inertial lengths $d_{i,e0}$ are defined in \citeA{CassakandShay.2007} and \citeA{Cassak.2017a}, where aforementioned ``hybrid'' quantities are intermediate in value between those of the two inflow regions.}
\begin{tabular}{| l | r | r | c | l | r |}
\cline{1-3}\cline{5-6}
& \multicolumn{1}{|c|}{Sphere} & \multicolumn{1}{|c|}{Sheath} & \hspace{3mm} &  \multicolumn{2}{|c|}{Boundary params.} \\
\cline{1-3}\cline{5-6}
\cline{1-3}\cline{5-6}
\multirow{3}{*}{$\left< \vec{B} \right>$ [nT]} & 58.12 $\hat{L}$ & --48.12 $\hat{L}$  & \cellcolor[gray]{0.8} & $|B_{L,sh}/B_{L,sp}|$ & 0.83 \\ 
\cline{2-3}\cline{5-6}
                                           &  7.02 $\hat{M}$ &  25.18 $\hat{M}$ & \cellcolor[gray]{0.8}& $|\left<B_M\right>/B_{L0}|$ & 0.29 \\
\cline{2-3}\cline{5-6}
                                           &   2.56 $\hat{N}$ & 4.73 $\hat{N}$      & \cellcolor[gray]{0.8}& $n_{sh}/n_{sp}$ & 28 \\
\cline{1-3}\cline{5-6}
$\left<|\vec{B}|\right>$ [nT] & $ 59.60 $ & $ 54.51 $    & \cellcolor[gray]{0.8}& Shear angle & 144$^\circ$  \\
\cline{1-3}\cline{5-6}
$\left<V_{i,L}\right>$ $(\left<V_{i,M}\right>)$ [km/s] & $\sim$0 & 11.4 (--153) & \cellcolor[gray]{0.8} &  $\left<V_{X\textrm{-}line,LMN}\right>$ [km/s] & [--157, --68, --23] \\
\cline{1-3}\cline{5-6}
$\left<n\right>$ [cm$^{-3}$] &  1.4  &  38.0  & \cellcolor[gray]{0.8}& $ B_{L0}$  [nT] & 54.13 \\
\cline{1-3}\cline{5-6}
$\left<T_e\right>$ [eV]       &   253  &   33.6  & \cellcolor[gray]{0.8}& $V_{Ai0}$  [km/s] & 249 \\
\cline{1-3}\cline{5-6}
$\left<T_i\right>$  [eV]       &  2800  &  221  & \cellcolor[gray]{0.8}& $d_{i0}$ ($d_{e0}$)$$ [km] & 49.3 (1.15) \\
\cline{1-3}\cline{5-6}
\end{tabular}
\end{table}

\section{Evidence for the ion diffusion region}

\begin{figure}
\noindent\includegraphics[width=39pc]{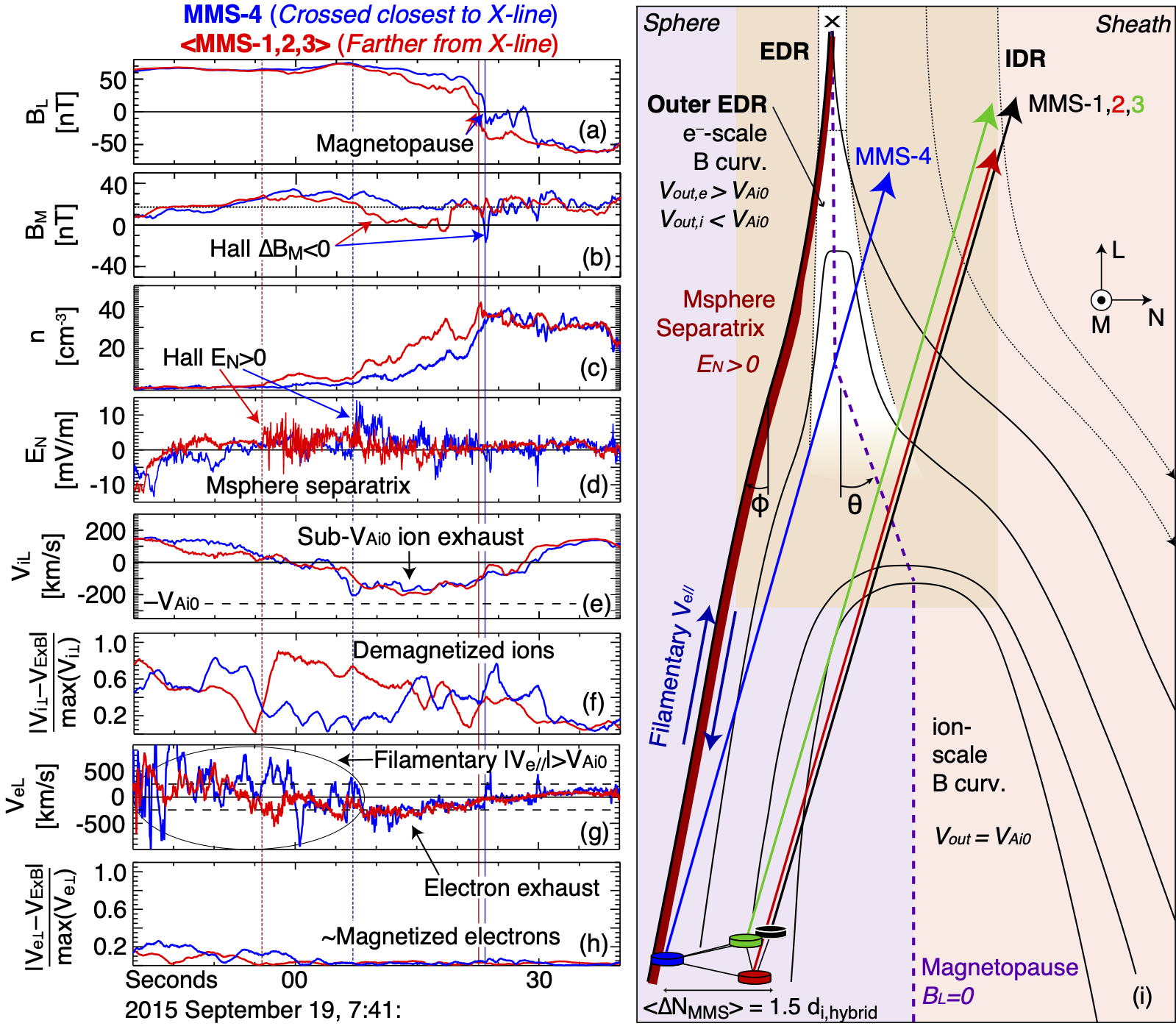}
\caption{(a): $B_L$, with the magnetopause crossings ($B_L$=0) labeled with vertical lines. (b): $B_M$. The asymptotic guide field is a horizontal dashed line. (c): Plasma number density. (d): $E_N$. The magnetospheric separatrix crossings are vertical lines. (e): $V_{iL}$. The upstream Alfv\'en speed is a horizontal dashed line. (f): The magnitude of the perpendicular ion bulk velocity in the frame of the convective electric field normalized by the maximum perpendicular ion speed. (g): $v_{eL}$. $\pm V_{Ai0}$ are horizontal dashed lines. (h): Similar to (f) but for electrons. (i): Schematic diagram of the magnetopause crossing, wherein only the relative separations between satellites are to scale. All data are in the co-moving reference frame. The red traces in (a)-(h) are the averaged quantities from MMS-1, 2, and 3.}
\label{ions}
\end{figure}

Many signatures of the IDR were observed, some of which are illustrated in Figure \ref{ions}i. Briefly, MMS observed filamentary super-ion-Alfv\'enic electron flows along the magnetosphere-side separatrix, sub-Alfv\'enic demagnetized ion outflow and magnetized electron outflow within the exhaust layer, and Hall electric and magnetic fields, all of which are shown in the Figure \ref{ions} in the co-moving velocity frame. As is depicted in Figure \ref{ions}i, MMS-4 crossed the magnetopause significantly ($\sim$100 km, $\sim$2 $d_{i0}$) nearer the X-line than MMS-1, 2, and 3, which had similar trajectories in the reconnection $L$-$N$ plane. As such, Figure \ref{ions}a-h compares MMS-4 with averaged data from the downstream spacecraft.

The magnetopause retreated inward across MMS, as shown by the transition from a low-density magnetospheric plasma with $B_L>0$ to a high-density magnetosheath plasma with $B_L<0$ (Fig. \ref{ions}a,c). Prior to crossing the magnetospheric separatrix, MMS observed filamentary field-aligned electron flows (Fig. \ref{ions}g), which have previously been reported downstream of magnetopause reconnection sites \cite{Phan.2016a,Genestreti.2018a,WangRS.2017}. The electron velocity remained positive on average but fluctuated in time, exhibiting large positive and negative values. These super-ion-Alfv\'enic electron flows adjacent to the separatrices carry the field-aligned portions of the Hall current loop. Upon crossing the separatrix, sunward Hall electric fields are observed (Fig. \ref{ions}d). The Hall $E_N>0$ is a finite ion gyroradius effect \cite{Pritchett.2008}. After crossing the separatrix, super-Alfv\'enic southward electron exhaust was observed (Fig. \ref{ions}g) along with sub-Alfv\'enic southward ion exhaust (Fig. \ref{ions}e). The ion exhaust was demagnetized, differing significantly (several hundred km/s) from the $\vec{E}\times\vec{B}$-drift velocity (Fig. \ref{ions}f). The electrons remained largely magnetized (Fig. \ref{ions}h). 

On the magnetospheric side of the exhaust, MMS-1, 2, and 3 observed a $\sim$10-second Hall out-of-plane field depression of $\Delta B_M\approx-21$ nT $\approx0.38B_{L0}$ (Fig. \ref{ions}b), where $\Delta B_M$ was calculated as the difference of $B_M$ and the guide field in the regions of the exhaust where the two differ most dramatically. MMS-4 observed a stronger ($\Delta B_M\approx35$ nT $\approx0.65B_{L0}$) Hall field at the magnetopause, which lasted $\sim$1 second. (Given the normal magnetopause speed of $-23$ km/s, the Hall $B_M$ region observed by MMS-4 was $\sim$23 km thick or $\sim20$ $d_{e0}$ or 0.5 $d_{i0}$. The Hall magnetic field is generated in the IDR as outflowing electrons drag the magnetic field in the out-of-plane direction \cite{Sonnerup.1979,Mandt.1994}. The Hall magnetic field region shifts toward the magnetospheric side of the exhaust with downstream distance as the perpendicular electron outflow is diverted by the dawn-ward guide field \cite{WangRS.2017}. Downstream of the IDR, the Hall magnetic field is maintained by the elongated field-aligned portions of the Hall current system \cite{Sonnerup.1979}. 

The strength and thickness of the Hall $\Delta B_M$ region are $\sim$70\% larger and $\sim$90\% narrower, respectively, on MMS-4 compared to the downstream spacecraft. The intense field aligned electron flows at the separatrix (maximum speed of 680 km/s $\approx2.6V_{Ai0}$ for MMS-4 versus 415 km/s $\approx1.6V_{Ai0}$ for the downstream spacecraft) were $\sim$65\% faster on MMS-4. The Hall electric field at the separatrix ($\sim$14 mV/m for MMS-4 versus $\sim$7 mV/m for the downstream spacecraft) was twice as large on MMS-4. 

Taken together, these observations indicate a crossing of the IDR dominated by Hall effects, which diminish over downstream distances on the order of two upstream ion inertial lengths.

\section{Opening angle of the exhaust within the IDR}

A lower-bound estimate of the opening angle of the exhaust within the IDR is determined as follows: (1) the magnetospheric separatrix normal is determined using timing analysis of the $E_N>0$ onset and MDD-B, (2) the magnetopause normal is determined by timing of $B_L=0$ and MDD-B, (3) the angle between the two normal directions in the reconnection plane is determined. In Figure \ref{ions}i, the angle of the magnetospheric separatrix is labeled $\phi$ and the angle of the magnetopause is labeled $\theta$. Both angles are defined relative to the $L$ axis. The angle $\theta+\phi$ is not necessarily equal to the half-angle of the exhaust in asymmetric reconnection \cite{Lee.2014,Liu.2018a}; regardless of the symmetry or asymmetry, and trivially, $\theta+\phi$ is always less than the total opening angle of the exhaust. 

The detailed calculations of $\phi$, $\theta$, and their uncertainties are found in Appendix B, the results of which demonstrate $\theta=24^\circ\pm4^\circ$, $\phi=6^\circ\pm4^\circ$, and $\theta+\phi=30^\circ\pm5^\circ$. The sum of $\theta+\phi$ is therefore dominated by $\theta$ by a 4-to-1 ratio. If the exhaust thickness increases monotonically and the co-moving velocity frame is valid for the whole crossing then MMS-4 entered the exhaust $\sim$60 $d_{i0}$ (2900 km) downstream of the X-point and crossed the magnetopause $\sim$7 $d_{i0}$ (380 km) downstream.

The geometry of the IDR is interconnected with the reconnection rate \cite{Liu.2017,Liu.2018a}. Following from Figures 3b and 3d of \citeA{Liu.2018a}, the normalized reconnection rate can be determined by the opening angles of either the magnetospheric or magnetosheath separatrix. The normalized rate is estimated to be $\approx$0.15 using $\phi=6^\circ$. Likewise, the expected angle of the magnetosheath separatrix is estimated to be $\sim20^\circ$. For comparison, we determined $\theta=24^\circ\pm4^\circ$. Given that the magnetosheath separatrix must have a larger angle than the magnetopause, the actual and predicted magnetosheath separatrix angles are therefore likely inconsistent, though the exact discrepancy is unknown. What is known is that the normalized reconnected magnetic flux observed at the magnetopause are highly structured and have maximum values that are significantly larger than 0.15 (note that, while this does not necessarily indicate a larger reconnection rate, it does indicate another discrepancy between predictions and MMS observations). The latter point is demonstrated and investigated in the remaining sections. Given that the theoretical framework of \citeA{Liu.2018a} was based on magnetohydrodynamic force balance, it is of particular interest to investigate whether non-ideal effects influence the IDR structure.

\section{Electron dynamics and flux pileup}

\begin{figure}
\noindent\includegraphics[width=32pc]{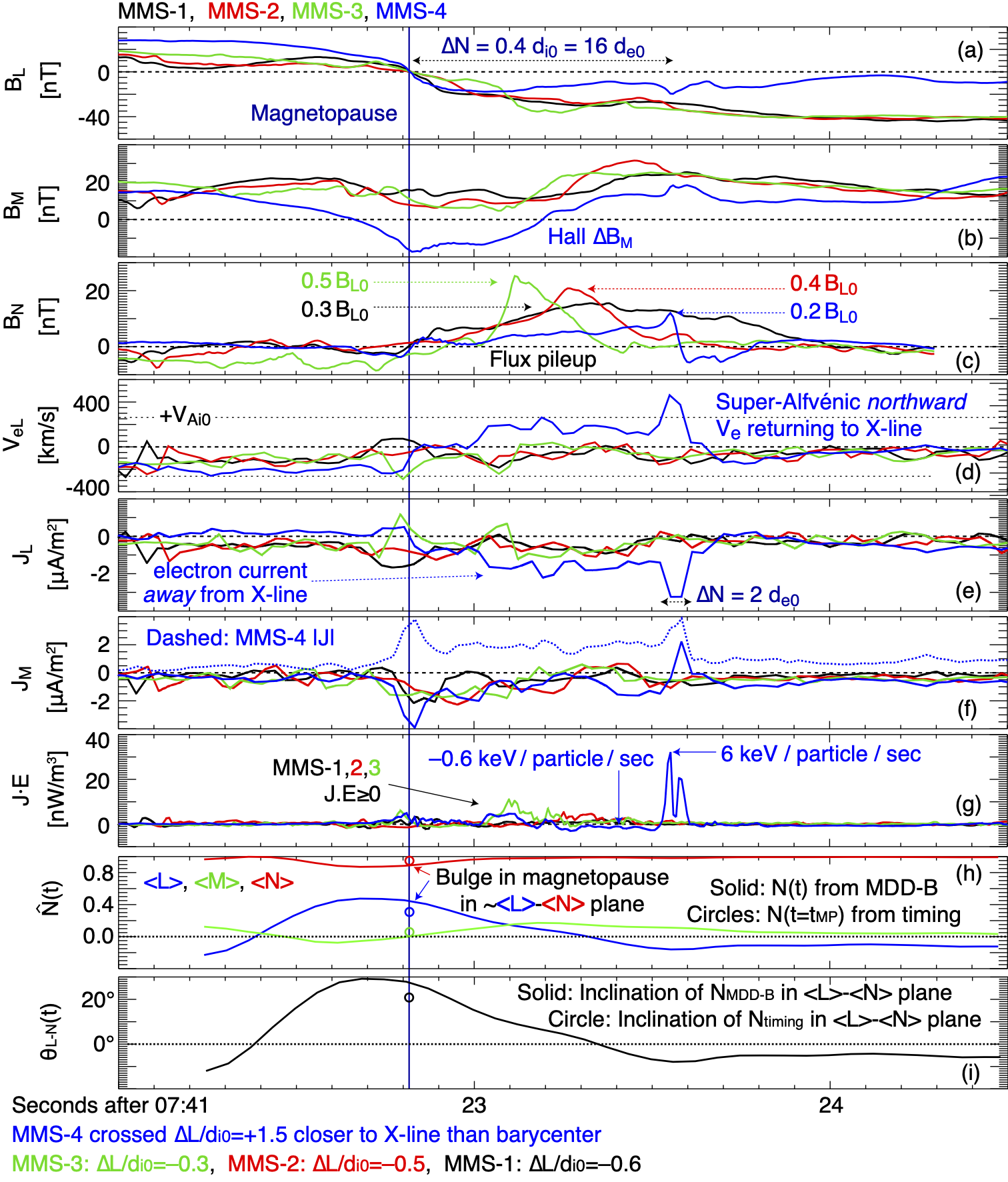}
\caption{(a), (b), and (c): The $L$, $M$, and $N$ components of the magnetic field. (d): The $L$ component of the electron bulk velocities. (e), (f): The $L$ and $M$ components of the current density calculated from the plasma moments. (g): The energy conversion rate $\vec{J}\cdot\vec{E}$. (h) The normal directions from the Maximum Directional Derivative of B (MDD-B) technique (solid line) and timing analysis (circles) shown in the L (blue) M (green) N (red) coordinates determined in appendix A and used throughout this study. (i) The angle of inclination of the normal directions from MDD-B (solid line) and timing analysis (circle) relative to the $L-N$ plane. In all but panels (h) and (i), black, red, green, and blue are used to represent MMS-1, 2, 3, and 4, respectively. All data are given in the co-moving reconnection velocity frame. The thickness of the pileup, determined as the duration times the magnetopause normal velocity, is labeled in (a). Data have been shifted in time to align the magnetopause crossings (blue vertical line); MMS-1 by +0.3 sec, MMS-2 by +0.03 sec, MMS-3 by +0.2 sec, and MMS-4 by --0.53 sec.}
\label{elecs}
\end{figure}

A number of small-scale features are observed within a second of the $B_L$ reversal (Figure \ref{elecs}), with scale sizes intermediate between the ion and electron inertial scales. Key to understanding these dynamics are the normal magnetic field (Fig. \ref{elecs}c), the electron outflow velocity (Fig. \ref{elecs}d), and the relative locations of the MMS spacecraft during the crossing (tabulated below Fig. \ref{elecs}); however, all data in Figure \ref{elecs} are required to understand the cause and effects of the electron-scale dynamics (see Section 7.2).

A channel of enhanced reconnected $B_N>0$ flux is observed by all spacecraft, the sign of which is expected southward of the X-line. The size of $B_N$ is almost half as large as the upstream field on MMS-3. The downstream spacecraft observed enhancements of $B_N\approx0.3-0.5B_{L0}$ (16-25 nT) extending from the magnetopause into the magnetosheath side of the boundary layer. The normalized $B_N$ is roughly consistent with the inclination of the magnetopause ($\mathrm{tan}^{-1}\theta=0.45$). The magnetopause is highly inclined relative to the overall $LMN$ coordinates, as shown in Fig. \ref{elecs}i and discussed in Appendix B. The inclination is principally in the $L-N$ plane, with the magnetopause normal being deflected northward along $L>0$ (Fig. \ref{elecs}h), as is drawn in Fig. \ref{ions}i. This feature is unique to asymmetric reconnection and typically occurs within 20 $d_{i0}$ of the X-line (see next section and Fig. 1 of \citeA{Shay.2016}, for instance). MMS-4 observed a smaller $B_N$ than the downstream spacecraft, with $B_N\approx0.2B_{L0}$ (11 nT). 

Unlike at the downstream spacecraft, the $B_N$ enhancement turns on and off sharply at MMS-4. The $B_N>0$ channel at MMS-4 is bounded on either side by sharp 2 $d_{e0}$-thick electron currents of almost 4 $\mu$A/m$^2$. The center of the $B_N>0$ channel at MMS-4 is observed in conjunction with a northward, i.e., flowing towards the X-line, flow (Fig. \ref{elecs}d) of magnetized electrons. The electrons flowing toward the X-line (observed by MMS-4) have oblique velocities relative to the magnetic field at the Earthward edge of the pileup and are anti-field-aligned at the magnetosheath-ward edge. The out-of-plane current was largely duskward, i.e., $J_M<0$, though some dawn-ward current is also observed in the vicinity of the $B_N>0$ maxima (Fig. \ref{elecs}f). Particle acceleration $\vec{J}\cdot\vec{E}>0$ was occurring at the downstream spacecraft, primarily at the Earthward edge of the flux pileup. MMS-4 observed very rapid $\vec{J}\cdot\vec{E}/n\approx$6 keV/particle/sec acceleration localized at the magnetosheath-ward edge of the $B_N>0$ channel and broader, slower particle deceleration within the channel (Fig. \ref{elecs}g). While MMS-4 observed mostly magnetized and gyrotropic (not pictured) northward electron flow, the downstream spacecraft observed weak and mostly southward electron outflow (Fig. \ref{elecs}d).


\section{Comparison with simulations}

We have performed two 2.5-dimensional fully-kinetic particle in cell (PIC) simulations to identify possible causes of the flux pileup. The details regarding the code and initial conditions are found in Appendix C. In brief, the first run (``Sim1'') has conditions closely matching the MMS data for our event, while the second run (``Sim2'') has a larger asymmetry in $B_L$. To compare our simulations with the MMS data, we consider the southward ($L<0$) exhaust near the X-point. All values are normalized by the relevant hybrid upstream values (e.g., $B_N$ is normalized by $B_{L0}$) to be consistent with the normalization scheme used thus far.

\begin{figure}
\noindent\includegraphics[width=28pc]{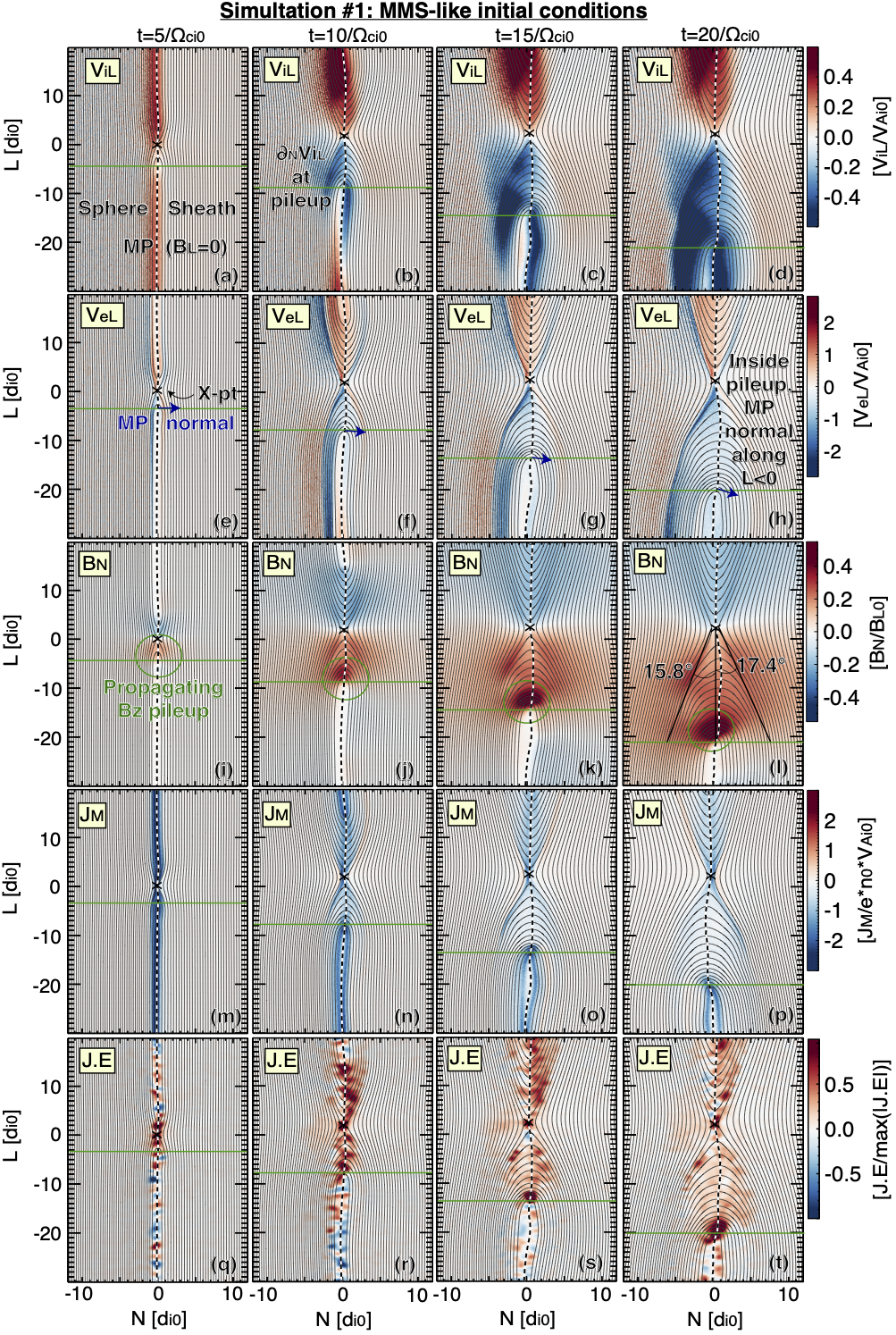}
\caption{Output of Sim1 at four times (first column) $t=$5 hybrid upstream ion cyclotron periods ($\Omega_{ci0}^{-1}$), (second column) $t=10\Omega_{ci0}^{-1}$, (third) $t=15\Omega_{ci0}^{-1}$, (fourth) $t=20\Omega_{ci0}^{-1}$. The thin black traces are the contours of $A_M$, the out-of-plane component of the magnetic vector potential. The thick white/black dashed lines are the magnetopause surfaces (where $B_L=0$). The vertical green lines track the front of the propagating $B_N$ pileup region. The color bars show normalized units, which are hybrid (combined from the two inflows) parameters, identical to those used in our MMS data analysis (c.f. Table 1) with the exception of panels q-t, $\vec{J}\cdot\vec{E}$, which are in (for the purposes of this article) arbitrary simulation units. In all panels, red indicates a positive quantity and blue indicates negative.}
\label{sim1}
\end{figure}

The output of Sim1 is shown in Figure \ref{sim1}. In Fig. \ref{sim1}i-l, there is a propagating flux pileup region associated with the rapid increase in the reconnection rate following the initiation of reconnection. This feature is likely analogous to dipolarization fronts in the magnetotail, as it is associated with the front of a jet that propagates rapidly through initially stationary ambient plasma. The flux pileup is observed on both sides of the magnetopause, which is weakly inclined relative to the simulation $L-N$ plane, where the magnetopause normal has a small component along $L<0$, i.e., away from the X-line and opposite to what was observed by MMS. The inclination of the magnetopause within the flux pileup at each of four times is drawn with dark blue arrows in Fig. \ref{sim1}e-h. The opening angles of the magnetospheric and magnetosheath separatrices ($15.8^\circ$ and $17.4^\circ$, respectively, as in Fig. \ref{sim1}l) are comparable to the expected values ($\sim17^\circ$ and $\sim19^\circ$, respectively) for steady reconnection with roughly symmetric $B_L$ and strongly asymmetric density \cite{Liu.2018a}; this indicates that the propagating flux pileup does not dramatically affect the exhaust opening angle, after its passage. Weak $\vec{J}\cdot\vec{E}<0$ may occur at the front's leading edge while strong $\vec{J}\cdot\vec{E}>0$ occurs in the trail (Fig \ref{sim1}q-t). The negative $\vec{J}\cdot\vec{E}$ is dominated by $J_LE_L$ (not pictured). Small-scale $\vec{J}\cdot\vec{E}$ may not be visible due to the smoothing applied to the electric field, which was done to filter noise. There is no apparent change in the electron bulk velocity near the pileup in its later stages (Fig. \ref{sim1}e-h); rather, the front is primarily associated with a sharp change in the ion bulk velocity (Fig. \ref{sim1}a-d). The strength of the pileup is weak when the pileup region exists near the X-point ($B_N\geq-0.18B_{L0}$) and becomes significantly larger than the background $B_N$ when the front propagates $\geq$10 $d_i$ downstream. The dimensions of the pileup are roughly 2 $d_{i0}$ in $L$ by 3 $d_{i0}$ in $N$ (note: $d_{i0}=\sqrt{75}d_{e0}$ in the simulation).

\begin{figure}
\noindent\includegraphics[width=18pc]{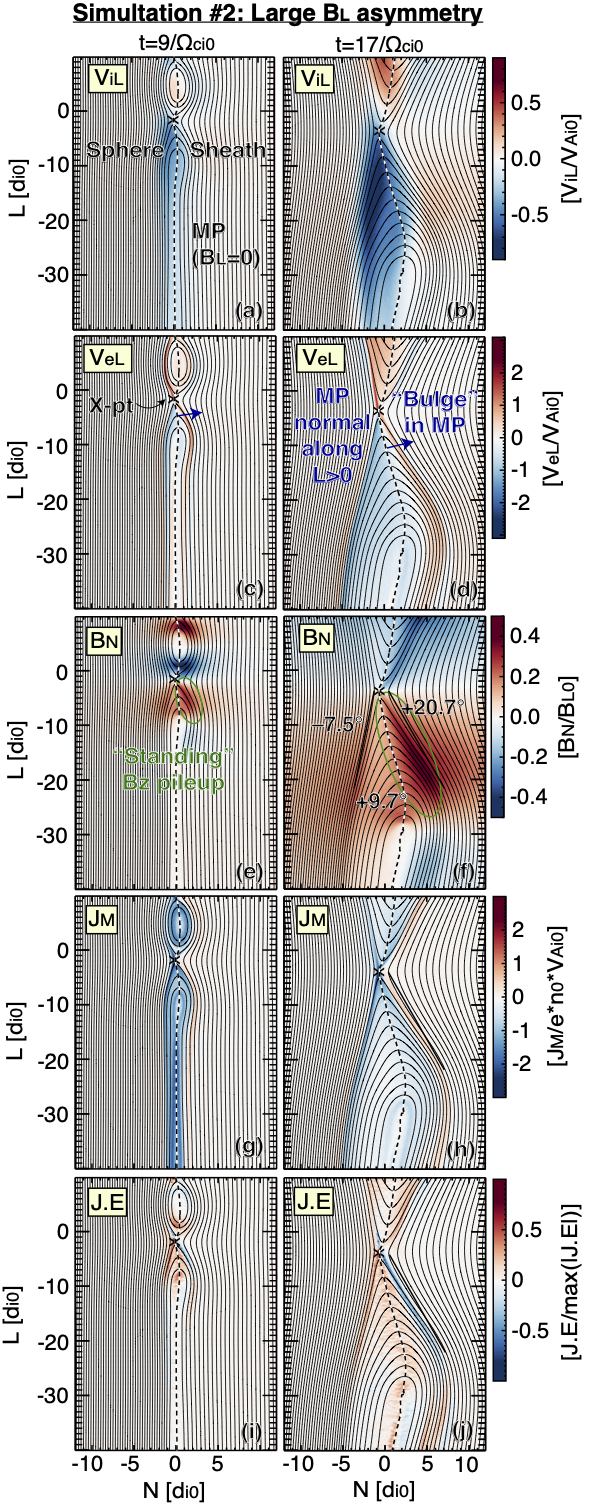}
\caption{Output of Sim2 at two times (first column) $t=9\Omega_{ci0}^{-1}$, (second) $t=17\Omega_{ci0}^{-1}$. The layout of rows in this Fig. is similar to Fig. \ref{sim1}.}
\label{sim2}
\end{figure}

The output of Sim2 is shown in Figure \ref{sim2}. In Fig. \ref{sim2}e-f, there is a standing $B_N$ pileup region that emanates from the X-point and lies between the magnetopause and the magnetosheath-side separatrix. The opening angles of the magnetosheath-side separatrix and magnetopause are both larger than that of the magnetosphere-side separatrix (Fig. \ref{sim2}f), though the angle between the magnetopause and the magnetosphere-side separatrix is larger than the angle between the magnetopause and the magnetosheath-side separatrix. The $B_N$ pileup is significantly larger than the background $B_N$; at time $t=17 \Omega_{ci0}^{-1}$, the largest value of $B_N\approx0.4B_{L0}$ (Fig. \ref{sim2}f). There is likely no analogous structure in symmetric reconnection since the feature appears to be associated with the stronger $B_L$ asymmetry used in Sim2. The existence of the large $B_N$ on the magnetosheath-side is not entirely unexpected, since, as mentioned earlier, it is easier for field lines to bend on the side of the exhaust that has smaller $B_L$. The $B_N$ pileup expands along $L$ as time progresses, but it does not disappear from the region very near the X-point as time progresses (compare Fig. \ref{sim2}e and \ref{sim2}f). The electron flow is positive (toward the X-line) on the magnetosheath-like side of the pileup and it reverses (flows away from the X-line) on the magnetosphere-like side. The magnetopause is inclined in the vicinity of the pileup, where the magnetopause normal is tilted $\sim10^\circ$ in the $L>0$ direction (Fig. \ref{sim2}f). The outflow component of the electron bulk velocity (Fig. \ref{sim2}c-d) is directed toward the X-point $V_{eL}>0$ at the magnetosheath-side edge of the pileup. At time $t=17 \Omega_{ci0}^{-1}$, within $\sim$15 $d_{i0}$ of the X-point, the electrons flowing toward the X-line near the sheath separatrix move obliquely relative the magnetic field, i.e., with both parallel and perpendicular velocity components (not pictured). Beyond $\sim15d_{i0}$ the separatrix electron flow is mostly anti-field-aligned. The energy conversion rate density $\vec{J}\cdot\vec{E}$ is less than zero within the portion of the pileup wherein the electrons are moving toward the X-point, while $\vec{J}\cdot\vec{E}\geq0$ elsewhere (Fig. \ref{sim2}i-j). The negative $\vec{J}\cdot\vec{E}$ is dominated by $J_LE_L$. At time $t=17 \Omega_{ci0}^{-1}$, the strongest negative $\vec{J}\cdot\vec{E}<0$ is observed within $\sim15$ $d_{i0}$ of the X-point. The dimensions of the pileup at this time are roughly 10 $d_{i0}$ along the magnetopause by 2 $d_{i0}$ perpendicular to the magnetopause. 

\section{Summary and interpretation of observations}

\subsection{Summary}

A primary magnetopause diffusion region was observed by MMS when the inter-probe separations were larger than the ion inertial scale. An ion-scale, sub-Alfv\'enic, demagnetized ion outflow was observed with an embedded super-ion-Alfv\'enic electron outflow. Field-aligned filamentary electron flows were observed at the magnetosphere-side separatrix and a normal Hall electric field was observed on the outflow side of the separatrix. The downstream spacecraft observed a broad ion-scale Hall magnetic field on the magnetospheric side of the outflow region, while MMS-4 (closest to the X-line) observed a narrow electron-scale Hall field at the magnetopause. The angle between the magnetopause and the magnetosphere-side separatrix was $\theta+\phi=30^\circ\pm5^\circ$. The magnetopause surface ($24^\circ\pm4^\circ$) was significantly more inclined than the magnetosphere-side separatrix ($6\pm4^\circ$) indicating the preferential sunward expansion of the reconnected field lines within the IDR, which displace the upstream magnetosheath. The large inclination angle of the magnetopause ($\mathrm{tan}^{-1}\theta\approx0.45$) coincided with a region of intense reconnected flux pileup with $B_N/B_{L0}$ of 0.3 to 0.5. Within the channel of piled-up $B_N$, MMS-4 observed a flow of mostly magnetized gyrotropic electrons flowing toward the X-line, with rapid particle acceleration occurring at the edges of the flow and deceleration occurring within. The pileup region was narrow ($\leq0.4$ $d_{i0}$ = 17 $d_{e0}$) in the inflow direction and long ($\geq$2 $d_{i0}$ = $87$ $d_{e0}$) in the outflow direction. The energy conversion rate density $\vec{J}\cdot\vec{E}$ observed by MMS-4 (nearest the X-line) was negative inside the pileup and strongly positive on the sheath-ward edge of the pileup. The downstream probes observed weak $\vec{J}\cdot\vec{E}\geq0$ in the vicinity of the pileup, with most of the energy conversion operating Earthward of the pileup.

Two simulations showed two unique scenarios with flux pileup. In simulation 1, a propagating flux pileup region was observed at the magnetopause at an ion jet front after reconnection was initiated. In simulation 2, a standing flux pileup region was observed that emanated from the X-point and lying between the magnetopause and the magnetosheath-side separatrix.

\subsection{Interpretation of results}

Simulation 1, which was based on MMS-like initial conditions, did not reproduce many key features observed by MMS, including the highly inclined magnetopause, including the location of the pileup (sheath-ward of the magnetopause), the dimensions of the pileup (very long in $L$ and very narrow in $N$), the strong reversals in $J_M$ and $V_{eL}$ within the pileup, etc. Simulation 2 reproduced these key features, at least qualitatively, though the strong $B_L$ asymmetry used in the simulation set up does not match the MMS observations in the asymptotic magnetosphere and magnetosheath. There was no indication from our analysis in Appendix A that the pileup region was moving relative to the overall magnetopause structure, which would rule out the propagating flow front scenario of simulation 1 (no clear jump in the Shi displacement when MMS moved into or out of the flux pileup in Fig. \ref{lmn_vstr}b-c); however, since the thickness of the pileup region ($\Delta N\sim0.4$ $d_{i0}$) was significantly smaller than the inter-probe separation ($\sim1.5$ $d_{i0}$), it is not clear whether our analysis techniques (the spatiotemporal difference (STD) technique and timing analysis) should be expected to correctly identify the speed of the pileup, were it propagating. We cannot conclusively distinguish between the scenarios of simulations 1 and 2, therefore, though we note that many aspects of the diffusion region suggest that it behaved as if there were a strong $B_L$ asymmetry. 

It is possible that the $B_L$ in the upstream magnetosheath varied before/during the crossing, such that the effective $B_L$ asymmetry was larger than what was determined in Table 1. It is also possible that the X-line was embedded within a region where $B_L$ at upstream of the diffusion region differed from the asymptotic $B_L$ in the magnetosheath and/or magnetosphere. However, without direct evidence for either of these scenarios, they remain purely speculative. Regardless of the cause, the reconnection diffusion region observed by MMS exhibited characteristics that are consistent with simulation 2, which had a strong $B_L$ asymmetry. We therefore consider simulation 2, and examine the force balance near the pileup to determine its cause.

\begin{figure}
\noindent\includegraphics[width=39pc]{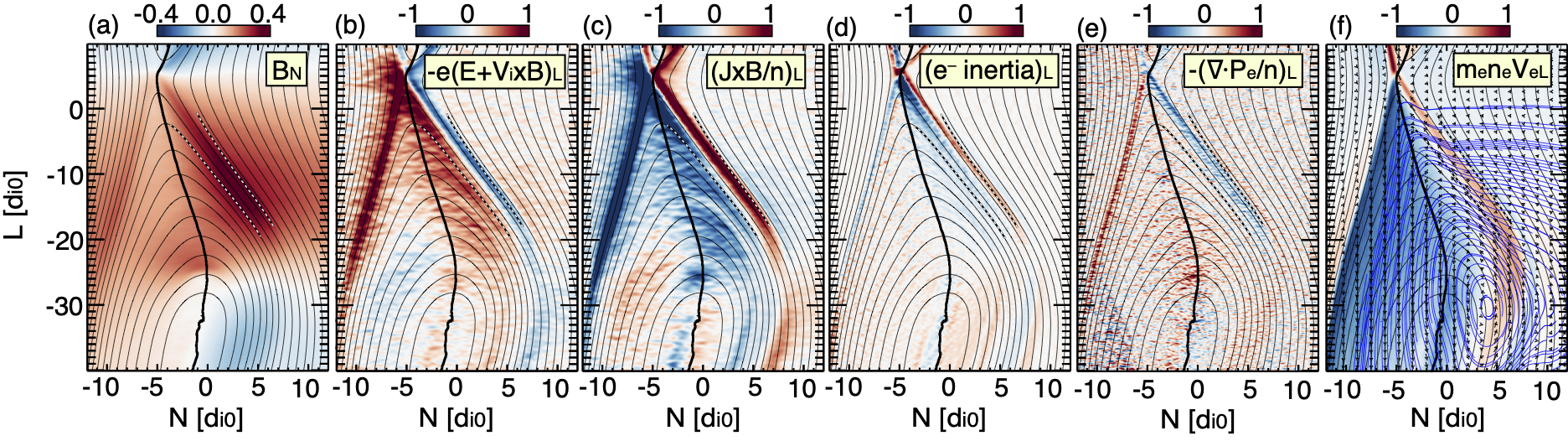}
\caption{Force balance terms near the pileup region in simulation 2 at $t=17\Omega_{ci0}^{-1}$. The units of the force terms are identical and arbitrary (normalized to the maximum value of $|e(\vec{E}+\vec{V}_i\times \vec{B})_L|$). }
\label{force}
\end{figure}

Figure \ref{force} shows key terms in the generalized Ohm's law from \citeA{Torbert.2016b} (neglecting the anomalous resistivity term and rewritten such that the electric field is in the ion frame), which is

\begin{equation}
-e\left(\vec{E}+\vec{V}_i\times\vec{B}\right)+\frac{\vec{J}\times\vec{B}}{n}+\frac{m_e}{n}\left(\frac{\partial\vec{J}}{e\partial t}+\nabla\cdot n\left(\vec{V}_i\vec{V}_i-\vec{V}_e\vec{V}_e\right)\right)-\frac{1}{en}\nabla\cdot\bar{P}_e=0
\end{equation}

\noindent where $e$ is the elementary charge. In simulation 2 and at MMS-4, the $\vec{J}\cdot\vec{E}$ associated with the pileup is dominated by $J_LE_L<0$. Figure \ref{force} specifically examines the force balance along the outflow ($L$) direction to determine the forces that balance $E_L$. The $E_L$ force term is shown in Figure \ref{force}b. At the magnetosheath-ward edge of the pileup (Fig. \ref{force}a) it points downward, away from the X-line (meaning the electric field points toward the X-line). On the magnetospheric side of the pileup, $-eE_L$ points toward the X-line. The Hall term is the dominant counterbalance of the electric force; however, since $\vec{J}\cdot(\vec{J}\times\vec{B})=0$, the Hall force can not exchange energy between the particles and the field. 

The candidates for balancing the force associated with the dissipative portion of the electric field are therefore the electron (Fig. \ref{force}d) and ion inertial forces and the electron pressure force (Fig. \ref{force}e). The ion inertial force is significantly weaker than all the other terms and is not pictured. The electron inertial force is weaker than the Hall force, but on either side of the pileup, it is directed oppositely to the electric force. Figure \ref{force}f illustrates the relevant behavior of the electron momentum (color) and velocity (arrows and streamlines). On the magnetosheath side, higher momentum electron inflow carries particles toward the X-line. Between the separatrix and the magnetopause, the electrons turn into the outflow direction, where they jet away from the X-line on the low-density magnetospheric side. The separatrix flows impart energy energy to the magnetic field as they brake and turn, generating the large $B_N$ (Fig. \ref{force}a). Because the magnetic field strength is weaker on this side, the large $B_N$ and the small $B_L$ causes the field to open at a wider angle on the higher density, lower $B_L$ magnetosheath side. The sum of all of the forces shown in Fig. 5 is small but non-zero, being largest (within $<10\%$ of the electric force) near the separatrices. The small remainder may be balanced by the time derivative term and/or it may result from our spatial smoothing and/or first-order differentiation.

\subsection{Conclusions}

We therefore make the following conclusion regarding the flux pileup observed by MMS: (1) the pileup, (2) the wider opening angle on the magnetosheath side of the exhaust, and (3) the flow of electrons toward the X-line observed by MMS-4 are all interconnected via the Hall effect. Effects (1) and (2) are unique to asymmetric reconnection, and they require asymmetric inflow conditions. The role of the $B_L$ asymmetry is yet unclear, since MMS did not show a clear and strong $B_L$ asymmetry. The exhaust opening angle is a critically important feature of magnetic reconnection, at least in part because it is thought to control the stability of the reconnection \cite{Liu.2015}. When the inflow magnetic fields are separated by a larger distance (e.g., when the opening angle is larger), they are less likely to filament and develop secondary reconnection sites. 

The overarching question of this investigation was: ``how, if at all, does bursty electron acceleration in the central EDR affect the ion-scale reconnection rate?'' We have provided evidence, for one diffusion region, that the Hall electron flows within the IDR play an important role in increasing the opening angle of the separatrices. If the perpendicular outflow portion of the Hall current is temporarily enhanced in the diffusion region, then one might expect the closure (inflow and outflow) Hall currents that generate the pileup to be similarly enhanced. In the future, it is desirable to examine more magnetopause ion diffusion regions with the MMS data in order to determine (1) whether the pileup is a regular feature, (2) whether the pileup is typically propagating or standing, and (3) whether the intensity of the pileup is related to bursts of electron acceleration in the EDR. While it is highly desirable to pursue this investigation, we note that MMS rarely used this type of large-scale configuration, though future extended mission operations may permit this type of investigation.

\appendix
\section{Orientation and motion of the reconnection region}

$LMN$ coordinates are determined with joint variance analysis (JVA) \cite{MozerandRetino.2007}, where maximum variance analysis (MVA) of $\vec{B}$ is used to determine $\hat{L}$, MVA of $\vec{E}$ is used to determine $\hat{N}$ that is adjusted to be perpendicular to $\hat{L}$, and $\hat{M}=\hat{N}\times\hat{L}$. JVA is applied to each spacecraft after smoothing $\vec{B}$ and $\vec{E}$ with a low-pass filter with a 3-second window, such that fluctuations that are unrelated to the overall boundary structure are removed. Ultimately, four $LMN$ coordinate systems are determined with JVA, whose $\hat{L}$, $\hat{M}$, and $\hat{N}$ axes differ from one another on average by $\sim2^\circ$. The system determined by applying JVA to MMS-4 data is used in this study as the eigenvalue separation was largest (the ratio of the $\hat{L}$ and $\hat{M}$ eigenvalues was 55.0 for MVA-$\vec{B}$ and the ratio of the $\hat{N}$ and $\hat{L}$ eigenvalues was 53.8 for MVA-$\vec{E}$). MVA is applied to the magnetic field data within the period 07:40:40--07:41:40 UT and to the electric field data within the period 07:40:30--07:41:40 UT. In geocentric solar ecliptic (GSE) coordinates, the axes are $\hat{L}=$[0.23178,  0.11340,  0.96613], $\hat{M}=$[0.50786, --0.86119, --0.02076], and $\hat{N}=$[0.82967,  0.49547, --0.25720].

\begin{figure*}
\noindent\includegraphics[width=30pc]{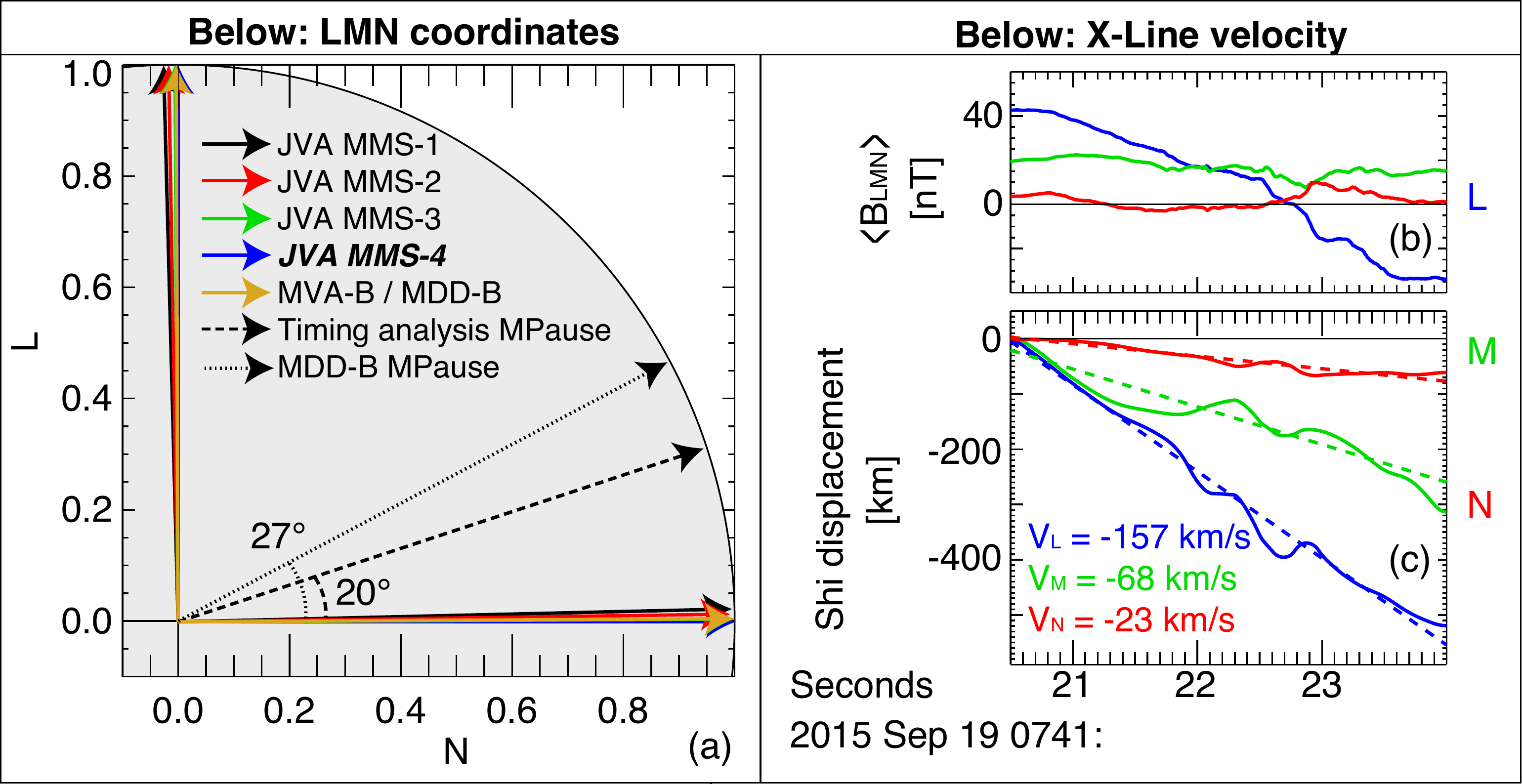}
\caption{(a) (solid arrows) the $LMN$ axes used in this study are determined with JVA applied to MMS-4, though as a quality check, four additional coordinate systems are determined with JVA and the MDD-B/MVA-B hybrid technique of \citeA{Denton.2018}. (Dashed arrow) the magnetopause surface normal determined with timing analysis. (Dotted arrow) the magnetopause surface normal determined with MDD-B at 07:41:22.8 UT.} (b) the average magnetic field vector for the 4 spacecraft and (c) the displacement and structure velocity determined with STD.
\label{lmn_vstr}
\end{figure*}

The results of JVA compare favorably with a fifth $LMN$ system, which is determined with the hybrid technique of \citeA{Denton.2018}, where the maximum directional derivative of $\vec{B}$ determines $\hat{N}$, MVA-$\vec{B}$ determines $\hat{L}$ that is adjusted to be perpendicular to $\hat{N}$, and $\hat{M}$ completes the right-handed system. The four sets of $LMN$ axes are all within $\leq4^\circ$ from those determined with JVA of MMS-4 data. The $\hat{L}$ and $\hat{N}$ axes of these five coordinate systems are compared in Figure \ref{lmn_vstr}a.

We apply four-point timing analysis \cite{ISSIchap10} to the $B_L$ reversal at the magnetopause. We obtain similar results for the magnetopause normal direction and speed that were determined in \citeA{WangS.2016} using timing analysis of $B_Z=0$. This normal direction of the $B_L=0$ surface is dramatically different from the previous normal directions, being tilted northward by $\sim20^\circ$. The normal direction obtained with timing analysis of the $B_L$ reversal is $\hat{N}_{MPause}=$[0.88809, 0.45605, 0.05748] (in GSE) and the speed of the magnetopause along the normal direction is $\vec{V}_{MPause}\cdot\hat{N}_{MPause}=-77$ km/s. For comparison, MDD-B is also used to determine the magnetopause surface normal. As shown in Fig. \ref{elecs}h, the MDD-B normal direction varies as a function of time in the vicinity of the magnetopause surface and flux pileup. However, at the time of the magnetopause surface crossing (07:41:22.8 UT, marked with a vertical blue line on Fig. \ref{elecs}), the deflection of the MDD-B is near its maximum, being tilted northward by $\sim28^\circ$. The two magnetopause surface normals, determined with MDD-B and timing analysis, are shown as dotted and dashed arrows, respectively, in Fig. \ref{lmn_vstr}a.

The full three-dimensional velocity of the reconnection layer is determined using the spatiotemporal difference method (STD) \cite{Shi.2006,Shi.2019}. STD is applied to the $\sim$3.5-second interval surrounding the magnetopause crossing, as shown in Figure \ref{lmn_vstr}b-c. STD assumes that the time variations of the magnetic field observed in the spacecraft reference frame are due to the advection of a steady-state structure. The reference frame of the structure $\vec{V}_{str}$ is found such that $\partial\vec{B}/\partial t=-\vec{V}_{str}\cdot\nabla\vec{B}$, where $\partial\vec{B}/\partial t$ is the time derivative in the spacecraft frame and a solution for $\vec{V}_{str}$ is most easily obtained in the eigenvector system of the 3x3 matrix $(\nabla\vec{B})^\mathrm{T}(\nabla\vec{B})$. It is possible to define a (time-dependent) solution for $\vec{V}_{str}$ for each magnetic field measurement made by MMS; however, we discard solutions that are associated with eigenvalues ten times smaller than the reported sensitivity of the MMS magnetometers, which is 0.1 nT \cite{Russell.2016}. The solutions for $\vec{V}_{str}$ that pass this quality criterion are integrated to obtain a displacement vector, which is then fit using linear regression (Figure \ref{lmn_vstr}c). The resulting velocity obtained with STD is, in $LMN$ coordinates, [--157, --68, --23] km/s. The projection of this velocity onto the normal vector obtained by timing analysis is --75 km/s, in good agreement with the previous estimate (--77 km/s).

\section{Opening angles and uncertainties of the magnetopause and separatrix}

The opening angle of the magnetopause was determined to be $\theta=20^\circ$ using timing analysis and $\theta=27^\circ$ using the maximum directional derivative of the magnetic field (MDD-B) technique (Fig. \ref{lmn_vstr}a). In asymmetric reconnection simulations, the magnetopause is similarly inclined within $10-20$ $d_{i0}$ of the X-line \cite{Shay.2016,Phan.2016b,WangS.2016}. The magnetopause $B_L=0$ surface is a sharp boundary, so uncertainties in the timing analysis resulting from timing are minimal and related to the small $\leq100$ pT uncertainties in the magnetic fields and the finite 128 Hz resolution of the magnetometers. Uncertainties related to the definition of $B_L=0$ stemming from the $LMN$ coordinates are likewise minimal; for instance, the magnetopause normal determined in each of the five coordinate systems given in appendix A differ from one another by less than 2$^\circ$. The 7$^\circ$ discrepancy between the magnetopause normal directions determined by timing analysis and MDD-B may indicate that (1) that the maximum gradient direction of $\vec{B}$ is not perpendicular to the $B_L=0$ surface and/or (2) that the magnetopause normal direction is not constant along the portion of its surface contained within the tetrahedron. Analysis in Section 6 reveals possible evidence for explanation (1); external to the magnetopause surface, strong $B_N$ pileup is observed with $\partial B_N/\partial L<0$ and MMS-4, uniquely, observed a large Hall-like deflection of $B_M$. Analysis in Fig. \ref{elecs}h-i reveals evidence for explanation (2); the MDD-B normal direction becomes more or less inclined relative to the overall $LMN$ coordinates of the boundary layer during the transit of MMS through the magnetopause, varying by $\pm7^\circ$ within the $\sim$half-second transit of the magnetopause by the four spacecraft. After factoring in our (somewhat crude) estimates for the uncertainties, we find a nominal estimate of $\theta=24^\circ\pm4^\circ$.

The opening angle of the magnetosphere-side separatrix $\phi$ was determined to be $5^\circ\pm3^\circ$ with timing analysis and 8$^\circ\pm4^\circ$ with MDD-B. Uncertainties in $\phi$ calculated with timing analysis are due to the fact that the large $E_N>0$ onset occurred over a finite range of time and thus does not define a `sharp' surface-like boundary. Uncertainties in $\phi$ calculated with MDD-B are likely due to the low current density, i.e., weak magnetic field gradient, which may not be fully resolved by MMS. Still, the MDD-B normal direction was steady and well-separated from the $L$ and $N$ directions (i.e., eigenvalue separations larger than 10) between 07:40:47–49 UT and 07:41:02–04 UT. The averaged normal direction obtained during these two intervals within the separatrix crossing is used to calculate $\phi=8^\circ\pm4^\circ$. The opening angle of the separatrix $\phi$ is taken to be the average of the values determined with timing analysis and MDD-B, i.e., $\phi=6^\circ\pm4^\circ$.

Uncertainties in the $L-N$ coordinates are likely comparable to all aforementioned uncertainties ($\pm2^\circ$ in Fig. \ref{lmn_vstr}a) but, while they may contribute uncertainty to either $\phi$ or $\theta$, they do not contribute significantly to the uncertainty in $\phi+\theta$. Combining our estimates for $\phi$, $\theta$, and their uncertainties yields $\phi+\theta=30^\circ\pm5^\circ$.

\section{Particle-in-cell (PIC) simulation set up}

This study uses the 2.5-dimensional VPIC code \cite{Bowers.2008} and we have performed two different runs. Both runs use an ion-to-electron mass ratio of $m_i/m_e=75$. The simulations have domain size $L\times N=102.4 d_i \times 51.2 d_i$, which is $4096\times2048$ cells. Periodic boundary conditions are used in the $L$ direction with conducting (reflecting) boundary conditions are used for fields (particles) in $N$. The subscripts ``sh'' and ``sp'' hereafter refer to the conditions of the magnetosheath $(N>0)$ and magnetosphere $(N<0)$ sides, respectively. The constants used in the simulation, which are typically given the subscript of ``0'' in other studies, are given the subscript of ``1'' below, such that they are not confused with the hybrid upstream values that have been thus far referred to with ``0''.

\subsection{Run 1: upstream conditions closely matching the event}
For the first simulation, which has initial conditions chosen to closely match the event parameters given in Table 1; especially the ratios of $B_{L,sh}/B_{L,sp}$ and $n_{sh}/n_{sp}$ and the strength of the guide field. The initial magnetic field configuration is $B_{L}(N)=B_{1}\left[0.5-\alpha_1\mathrm{tanh}(N/\delta)\right]$ and $B_M=B_1b_g$. The initial density is $n(N)=\alpha_3n_1\left[1+\alpha_2\mathrm{tanh}(N/\delta)\right]$ and the temperature is $T=T_i+T_e=\left[\alpha_4-B_L^2/8\pi\right]/n$. Inside the current sheet, the ion-to-electron drift speed ratio is $T_i/T_e$. Background shear flows are added such that $V_L(N)=-\alpha_5\mathrm{tanh}(N/\delta)$ and $V_M(N)=-\alpha_6\mathrm{tanh}(N/\delta)$ for both ions and electrons. The fields are then rotated $9.7^\circ$ about the $N$ direction such that the $M$ direction nearly bisects the upstream magnetic fields, as was the case for the MMS event. To match the event, we used $b_g=-3.771$, $\alpha_1=12.155$, $\alpha_2=0.81$, $\alpha_3=0.553$, $\alpha_4=0.035$, $\alpha_5=0.109V_{A,sh}$, and $\alpha_6=0.437V_{A,sh}$. The initial half-thickness is $\delta=0.5d_i$, where the ion inertial length is $d_i=c/(4\pi n_1e^2/m_i)^{1/2}$. The ion-to-electron temperature ratio is $T_i/T_e=10$. The ratio of the electron plasma to gyro frequencies is $\omega_{pe}/\Omega_{ci}=4$, where $\omega_{pe}=(4\pi n_{1} e^2/m_e)^{1/2}$ and $\Omega_{ci}=eB_{1}/m_ec$.

\subsection{Run 2: generic upstream conditions with $B_L$ asymmetry}
For the second simulation, which has initial conditions matching those of \cite{Liu.2018b} but with a larger mass ratio. The asymmetric current sheet has an initial magnetic field of $B_L(N)=B_1(0.5-S(N))$, where $S(N)=\mathrm{tanh}(N/\delta)$, and $B_M=-B_1$. The profile of $B_L$ gives upstream values of $B_{L,sh}=-0.5B_1$ and $B_{L,sp}=1.5B_1$. The initial half-thickness is $\delta=0.5d_i$. The initial density is $n(N)=n_1[1-(S^2-S)/3]$, corresponding to $n_{sh}/n_{sp}=3$.The temperature is uniform and the temperature ratio is $T_i/T_e$=5. Inside the current sheet, the particles drift with $-V_{di}/V_{de}=T_i/T_e$. The ratio of the electron plasma to gyro frequencies is $\omega_{pe}/\Omega_{ci}=4$.

\acknowledgments
We would like to acknowledge the efforts of the many MMS team members that made the mission successful. This study has employed several routines from the Space Physics Environment Data Analysis System \cite{spedas} and has benefited from discussions with Dr. Paul Cassak, Dr. Charlie Farrugia, Dr. Terry Forbes, Tony Rogers, and Dominic Payne. MMS data are available publicly from the mission's Science Data Center (https://lasp.colorado.edu/mms/sdc/public/), with the exception of the level 3 electric field data, that are available from the EDP team upon request. The solar wind data used in this study are from the WIND mission and are available publicly from CDAWeb (https://cdaweb.sci.gsfc.nasa.gov). Simulation outputs are available upon request to KJG. KJG was supported by NASA's MMS FIELDS grant NNG04EB99C. RED was supported by NASA grant 80NSSC19K0254. Research at LASP was supported by NASA grant NNX14AF71G. T.D.P was supported by NASA Grant 80NSSC18K0157. YHL was supported by NASA MMS 80NSSC18K0289.


%

%




\end{document}